\documentstyle[preprint,eqsecnum,aps]{revtex}
 \tightenlines
\newcommand{\be}{\begin{equation}}
\newcommand{\ee}{\end{equation}}
\newcommand{\bea}{\begin{eqnarray}}
\newcommand{\eea}{\end{eqnarray}}
\newcommand{\ba}{\begin{array}}
\newcommand{\ea}{\end{array}}

\newcommand{\Th}{\Theta}

\newcommand{\Ga}{\Gamma}

\newcommand{\La}{\Lambda}
\newcommand{\la}{\lambda}
\newcommand{\lb}{\label}

\newcommand{\de}{\delta}

\newcommand{\pa}{\partial}

\newcommand{\no}{\nonumber}

\newcommand{\tr}{\mbox{tr}}
\newcommand{\res}{\mbox{res}}

\begin{document}
\draft
\title
{\sc Poisson Algebras associated with Constrained
\\Dispersionless Modified KP Hierarchies\/}
\author{
{\sc Jen-Hsu Chang\/}
 \footnote{E-mail: changjen@math.sinica.edu.tw}\\
  {\it Institute of Mathematics, Academia Sinica,\\
   Nankang, Taipei, Taiwan\/}\\
    and\\
 {\sc Ming-Hsien Tu\/}\footnote{E-mail:
 phymhtu@ccunix.ccu.edu.tw}\\
  {\it Department of Physics, National Chung Cheng University,\\
   Minghsiung, Chiayi, Taiwan\/}
}
\date{\today}
\maketitle
\begin{abstract}
We investigate the bi-Hamiltonian structures associated with
constrained dispersionless modified KP hierarchies which are
constructed from truncations of the Lax operator of the
dispersionless modified KP hierarchy. After transforming their
second Hamiltonian structures to those of Gelfand-Dickey type, we
obtain the Poisson algebras of the coefficient functions of the
truncated Lax operators. Then we study the conformal property and
free-field realizations of these Poisson algebras.  Some examples
are worked out explicitly to illustrate the obtained results.
\end{abstract}
 \pacs{ }

\newpage


\section{Introduction}

The dispersionless integrable hierarchies can be viewed as the
quasi-classical limit of the ordinary integrable systems
\cite{LM}. A typical example is the dispersionless
Kadomtsev-Petviashvili (dKP) hierarchy which has played an
important role in theoretical and mathematical physics (see, for
example, \cite{TT} and references therein). The Lax formulation of
the dKP hierarchy can be constructed by replacing the
pseudo-differential Lax operator of KP with the corresponding
Laurent series. On the other hand, an analogue construction can be
made for the modified KP (mKP) hierarchy and thus leads to the
dmKP hierarchy.

In the previous work \cite{CT}, we established the Miura map
between the dKP and dmKP hierarchies, which turns out to be
canonical in the sense that the bi-Hamiltonian structure of the
dmKP hierarchy \cite{Li} is mapped to the bi-Hamiltonian of the
dKP hierarchy \cite{Li,FR}. We also studied the solution structure
of the dmKP hierarchy using the twistor construction \cite{TT}. In
this paper we turn to the Poisson algebras of the bi-Hamiltonian
structures associated with the dmKP hierarchy and, particularly,
its reductions. For the ordinary mKP hierarchy the reductions are
quite limited \cite{KO,OS,O}. However, in the dispersionless
limit, we show that the Lax operator of the dmKP hierarchy can be
truncated to any finite order and their associated bi-Hamiltonian
structures can be obtained via the Dirac reduction. To proceed the
formulation of the dmKP hierarchy, we recall some basic facts
about the algebra of Laurent series in the following.

 Let $\La$ be an algebra of Laurent series of the form
 \[
\La=\{A|A=\sum_{i=-\infty}^Na_ip^i\},
 \]
  with coefficients $a_i$
depending on an infinite set of variables $t_1
 \equiv x, t_2, t_3, \cdots$.
The algebra $\La$ can be decomposed into the subalgebras as
 \[
\La=\La_{\ge k}\oplus \La_{< k},
 \]
 where
  \bea \La_{\ge
k}&=&\{A\in \La| A=\sum_{i\ge k}a_ip^i\}\no\\
 \La_{< k}&=&\{A\in\La| A=\sum_{i< k}a_ip^i\}\no
 \eea
 and using the notations :
$\La_+=\La_{\ge 0}$ and $\La_-=\La_{< 0}$ for short. Although
$\La$ form a commutative and associative  algebra under
multiplication, we can define a Lie-bracket associated with $\La$
such that
 \[
\left[\left[A, B\right]\right]=\frac{\pa A}{\pa p}\frac{\pa B}{\pa
x}-\frac{\pa A}{\pa x}\frac{\pa B}{\pa p}, \qquad A,B \in \La
 \]
which can be regarded as the Poisson bracket defined in the
2-dimensional phase space $(x, p)$. For a given Laurent series $A$
we define its residue as
 \[
\res A=a_{-1}
 \]
and its trace as
 \[
\tr A=\int \res A.
 \]
 For any two Laurent series $A=\sum_ia_ip^i$ and
$B=\sum_ib_ip^i$ we have
 \[ \res \left[\left[A,
B\right]\right]=\sum_i i(a_ib_{-i})'
 \]
 which implies
 \[
\tr \left[\left[A, B\right]\right]=0,
 \]
and
 \[
\tr (A\left[\left[B, C\right]\right])=\tr (\left[\left[A,
B\right]\right]C).
 \]
  Finally, given a functional $F(A)=\int
f(a)$ we define its gradient as
 \[
\frac{\de F}{\de A}=\sum_i\frac{\de f}{\de a_i}p^{-i-1}
 \]
where the variational derivative is defined by
 \[
\frac{\de f}{\de a_k}=\sum_i(-1)^i\left(\pa^i\frac{\pa f}{\pa
a_k^{(i)}}\right),
 \]
with $a_k^{(i)}\equiv (\pa^ia_k), \pa\equiv \pa/\pa x$.

This paper is organized as follows: In section II, we will derive
the bi-Hamiltonian structures of constrained dmKP hierarchies from
the one of the dmKP hierarchy by the Dirac reduction. In section
III, we will investigate the conformal property of the second
Poisson brackets associated with constrained dmKP hierarchies. In
section IV, the free-field realizations of these Poisson algebras
will be given through the corresponding Kupershmidt-Wilson (KW)
theorem. We will give some examples to illustrate the obtained
results in section V. Section VI contains some concluding remarks.

\section{bi-Hamiltonian structures}

The (generalized) dmKP hierarchy is defined by the Lax operator of
the form
 \[\label{laxop}
  K_n=p^n+v_{n-1}p^{n-1}+\cdots,\qquad
(n>0)
 \]
  which satisfies the equations of motion
 \be\lb{lax}
\frac{d K_n}{dt_k}=\left[\left[B_k, K_n\right]\right],\qquad
B_k=(K_n^{k/n})_{\geq 1}
 \ee
 or zero-curvature conditions
 \be\label{zs}
 \frac{\pa B_k}{\pa t_l}-\frac{\pa B_l}{\pa t_k}+\left[\left[B_k, B_l\right]\right]=0.
 \ee
  For $n=1$, the first nontrivial flows ($t_2=y, t_3=t$) of (\ref{zs}) are given by
 \bea
 v_{-1x}&=&\frac{3}{2}v_{0y}-\frac{3}{2}(v_0^2)_x,\no\\
 v_{-1y}&=&2v_{0t}-\frac{3}{2}(v_0^2)_y-2v_{-1}v_{0x}.\no
 \eea
 which, by eliminating $v_{-1}$, yields the dmKP
equation
 \[
 v_{0t}=-\frac{3}{2}v_0^2v_{ox}+\frac{3}{2}v_{0x}\pa_x^{-1}v_{0y}+\frac{3}{4}\pa_x^{-1}v_{0yy}.
 \]
  The Hamiltonian structures associated with $K_n$ have been
obtained by Li \cite{Li} using the classical $r$-matrix
formulation. Especially, the second structure can be expressed as
 \[
\{F, G\}(K_n)=\int \res \left(J_2^{(n)}\left(\frac{\de F}{\de
K_n}\right)\frac{\de G}{\de K_n}\right)
 \]
  where the Hamiltonian map $J_2^{(n)}$ is defined by
 \be\lb{ham2}
J_2^{(n)}(X)=\left[\left[K_n, X\right]\right]_{\geq
-1}K_n-\left[\left[K_n, (K_nX)_{\geq 1}\right]\right]
 \ee
 with $X=\sum_ix_ip^{-i-1}$. It is quite natural to define the conserved
quantities as
 \[\label{conserve}
H_k=\frac{n}{k}\tr K_n^{k/n},
 \]
 then the Lax flows (\ref{lax}) can be
described by the Hamiltonian equations
 \[\lb{ham}
\frac{dK_n}{dt_k}=\{H_k,
K_n\}_2^{(n)}(K_n)=J_2^{(n)}\left(\frac{\de H_k}{\de K_n}\right).
 \]
Based on the above results, we would like to consider reductions
of the dmKP hierarchy and their associated Hamiltonian structures.
Let us consider truncations of the Lax operator $K_n$ as follows
 \be\label{mlax}
K_{(n,m)}=p^n+v_{n-1}p^{n-1}+\cdots+v_{-m}p^{-m},\qquad m\in
Z/\{0\}.
 \ee
 It is quite easy to show that these are consistent
truncations with respect to the Lax flows (\ref{lax}). Thus for
each pair $(n,m)$, the Lax operator $K_n$ with infinitely many
coefficient functions is reduced to a finite-dimensional one
$-K_{(n,m)}$ which we refer to the constrained dmKP hierarchy.
 However, the Hamiltonian map (\ref{ham2}) can not preserve the
form of $dK_{(n,m)}/dt_k$ since the lowest order term of
$J_2^{(n)}(\de H/\de K_{(n,m)})$ in $p$ is $p^{-m-1}$. Hence we
shall consider the Lax operator $\bar{K}_{(n,m)}=K_{(n,m)}+\mu
p^{-m-1}$ and then impose the constraint $\mu=0$ by the Dirac
reduction. It turns out that Hamiltonian flows for
$\bar{K}_{(n,m)}$ under the condition $\mu=0$ gives the second
class constraint:
 \be\lb{constraint}
\left(\res \left[\left[\bar{K}_{(n,m)}, \frac{\de H}{\de
\bar{K}_{(n,m)}}\right]\right]\right)_{\mu=0}=0
 \ee
  where
 \[
\frac{\de H}{\de \bar{K}_{(n,m)}}=\frac{\de H}{\de
K_{(n,m)}}+\frac{\de H}{\de \mu}p^{m}.
 \]
 That means the function
$\de H/\de \mu$ should be in terms of $\de H/\de v_i, i=-m,
-m+1,\cdots, n-1$. Solving the constraint (\ref{constraint}), we
obtain
\[
v_{-m}\frac{\de H}{\de \mu}=\frac{1}{m}\int^x
\res\left[\left[K_{(n,m)}, \frac{\de H}{\de
K_{(n,m)}}\right]\right]
\]
which implies
 \bea\lb{mham2}
  J_2^{(n)}\left(\frac{\de H}{\de \bar{K}_{(n,m)}}\right)&=&
\left[\left[K_{(n,m)}, \frac{\de H}{\de
\bar{K}_{(n,m)}}\right]\right]_{\geq -1}K_{(n,m)}-
\left[\left[K_{(n,m)}, \left(K_{(n,m)}\frac{\de H}{\de
\bar{K}_{(n,m)}}\right)_{\geq 1}\right]\right],\no\\
&=&\left[\left[K_{(n,m)}, \frac{\de H}{\de
K_{(n,m)}}\right]\right]_{\geq -1}K_{(n,m)}-
\left[\left[K_{(n,m)}, \left(K_{(n,m)}\frac{\de H}{\de
K_{(n,m)}}\right)_{\geq 1}\right]\right]\no\\
&&+\left[\left[K_{(n,m)}, v_{-m}\frac{\de H}{\de
\mu}\right]\right],\no\\ &=&\left[\left[K_{(n,m)}, \frac{\de
H}{\de K_{(n,m)}}\right]\right]_+K_{(n,m)}- \left[\left[K_{(n,m)},
\left(K_{(n,m)}\frac{\de H}{\de
K_{(n,m)}}\right)_+\right]\right]\no\\ &&+\left[\left[K_{(n,m)},
\left(K_{(n,m)}\frac{\de H}{\de
K_{(n,m)}}\right)_0\right]\right]+\left[\left[K_{(n,m)}, \frac{\de
H}{\de K_{(n,m)}}\right]\right]_{-1}K_{(n,m)}\no\\
&&+\frac{1}{m}\left[\left[K_{(n,m)},
\int^x\res\left[\left[K_{(n,m)}, \frac{\de H}{\de
K_{(n,m)}}\right]\right]\right]\right],\no\\ &\equiv &
J_2^{(n,m)}\left(\frac{\de H}{\de K_{(n,m)}}\right).
 \eea
 We note that the above modified Hamiltonian map
for $m=\pm 1$ are just the classical limit of the second
structures of mKP hierarchies obtained in \cite{KO,OS,O}. Besides,
when $m\to \infty$, $J_2^{(n,\infty)}$ recovers the Hamiltonian
map $J_2^{(n)}$, as expected.

Finally, we would like to remark that the first Poisson structure
of the constrained dmKP hierarchies can be defined as a
deformation of $J_2^{(n,m)}$ by shifting $K_{(n,m)}\mapsto
K_{(n,m)}+\lambda$ by a constant parameter $\lambda$. Then the
second structure induces a linear term $J_2^{(n,m)}\mapsto
J_2^{(n,m)}+\lambda J_1^{(n,m)}$ with
\begin{equation}\label{ham1}
  J_1^{(n,m)}\left(\frac{\de H}{\de K_{(n,m)}}\right)=
  \left[\left[K_{(n,m)},\frac{\de H}{\de K_{(n,m)}}\right]\right]_{\geq -1}-
  \left[\left[K_{(n,m)}, \left(\frac{\de H}{\de K_{(n,m)}}\right)_{\geq 1}\right]\right]
\end{equation}
which, by definition, is compatible with the second structure and
is a Laurent series of order at most $n-1$. It turns out that
(\ref{ham1}) is nothing but the first Poisson structure defined in
\cite{Li}. Note that the Hamiltonian map $J_1^{(n,m)}$ is
consistent with the Lax flows (\ref{lax}) for $m>0$ but not for
$m<0$ due to the fact that the lowest order term of
$J_1^{(n,m)}(\de H/\de K_{(n,m)})$ in $p$ is $p^{-1}$, not
$p^{|m|}$. Hence, just like the second structure, we require the
use of Dirac's theory of constraints to obtain the consistent
result. This will be done in the next section.


\section{Poisson algebras}

Having constructed the Hamiltonian map of the constrained dmKP
hierarchies, we are now ready to calculate the Poisson brackets of
the coefficient functions $v_i$ in (\ref{mlax}). Before doing
that, we would like to show that the complicated form of the
modified Hamiltonian map $J_2^{(n,m)}$ defined by $K_{(n,m)}$ can
be transformed to the familiar Gelfand-Dickey (GD) type structures
via the following identification
 \[
  K_{(n,m)}=L_{n+m}p^{-m}
=(p^{n+m}+u_{n+m-1}p^{n+m-1}+\cdots+u_0)p^{-m}
 \]
  where
 \be\lb{iden}
v_i=u_{i+m},\qquad i=-m, -m+1,\cdots, n-1. \ee On the other hand,
from the variation
 \[
\de F=\int \res \left(\de K_{(n,m)}\frac{\de F}{\de
K_{(n,m)}}\right)= \int \res \left(\de L_{n+m}\frac{\de F}{\de
L_{n+m}}\right)
 \]
  we have
 \[
\frac{\de F}{\de K_{(n,m)}}=p^m\frac{\de F}{\de L_{n+m}}.
 \]
Using the above relations, some terms in (\ref{mham2}) can be
rewritten as
 \bea
  \left[\left[K_{(n,m)}, \frac{\de F}{\de
K_{(n,m)}}\right]\right]_+K_{(n,m)} &=&
p^{-m}\left[\left[L_{n+m},\frac{\de F}{\de
L_{n+m}}\right]\right]_+L_{n+m}\no\\ &&-mK_{(n,m)}p^{-1}
\left(K_{(n,m)}\frac{\de F}{\de K_{(n,m)}}\right)'_+\no\\
&&-mK_{(n,m)}p^{-1}\left(K_{(n,m)}\frac{\de F}{\de
K_{(n,m)}}\right)'_0,\no\\
 \left[\left[K_{(n,m)}, \left(K_{(n,m)}\frac{\de F}{\de K_{(n,m)}}\right)_+\right]\right] &=&
p^{-m}\left[\left[L_{n+m}, \left(L_{n+m}\frac{\de F}{\de
L_{n+m}}\right)_+\right]\right] \no\\
&&-mp^{-1}K_{(n,m)}\left(K_{(n,m)}\frac{\de F}{\de
K_{(n,m)}}\right)'_+,\no\\
  \frac{1}{m}\left[\left[K_{(n,m)},\int^x\res \left[\left[K_{(n,m)},\frac{\de F}{\de K_{(n,m)}}\right]\right]\right]\right] &=&
\frac{1}{m}p^{-m}\left[\left[L_{n+m},\int^x\res\left[\left[L_{n+m},\frac{\de
F}{\de L_{n+m}}\right]\right]\right]\right]\no\\
&&-p^{-1}K_{(n,m)}\res\left[\left[K_{(n,m)},\frac{\de F}{\de
K_{(n,m)}}\right]\right]\no\\ &&-\left[\left[K_{(n,m)},
\left(K_{(n,m)}\frac{\de F}{\de
K_{(n,m)}}\right)_0\right]\right]\no\\
&&-mp^{-1}K_{(n,m)}\left(K_{(n,m)}\frac{\de F}{\de
K_{(n,m)}}\right)'_0,\no
 \eea
  which imply
  \bea\label{dictionary}
\{F,G\}_2^{(n,m)}(K_{(n,m)})&=&\int\res\left(J_2^{(n,m)}\left(\frac{\de
F}{\de K_{(n,m)}}\right)\frac{\de G}{\de K_{(n,m)}}\right),\no\\
&=&\int \res\left(\Th^{GD}_{2+3}\left(\frac{\de F}{\de
L_{n+m}}\right)\frac{\de G}{\de L_{n+m}}\right),\no\\ &=&\{F,
G\}^{GD}_{2+3}(L_{n+m})
 \eea
 where the Hamiltonian map $ \Th^{GD}_{2+3} \equiv \Th^{GD}_2+\frac{1}{m}\Th^{GD}_3 $
with
 \bea
\Th^{GD}_2(X)&=&\left[\left[L_{n+m},X\right]\right]_+L_{n+m}-\left[\left[L_{n+m},(L_{n+m}X)_+\right]\right],
\lb{gd2}\\
 \Th^{GD}_3(X) &=&\left[\left[L_{n+m}, \int^x\res\left[\left[L_{n+m},X\right]\right]\right]\right].\lb{gd3}
 \eea
Besides the standard second GD structure $\Th^{GD}_2$, (\ref{gd3})
is called the third GD bracket which is compatible with the second
one. Hence, under the identification (\ref{iden}), the modified
Hamiltonian structure (\ref{mham2}) has been mapped to the sum of
the second and the third GD   structures defined by the polynomial
$L_{n+m}$.

Since the Poisson algebras associated with the second GD structure
have been obtained \cite{FR}, we only need to treat the third one.
Therefore, by (\ref{dictionary}), we can now use (\ref{gd2}) and
(\ref{gd3}) instead of (\ref{mham2}) to read off the Poisson
brackets $\{v_i(x),
v_j(y)\}_2^{(n,m)}=-\left(J_2^{(n,m)}(v)\right)_{ij}\de(x-y)$
where the operators $\left(J_2^{(n,m)}(v)\right)_{ij}$ are taken
at the point $x$. After some straightforward algebras we have
 \bea\label{poal2}
\left(J_2^{(n,m)}\right)_{n-1,n-1}&=&\frac{n(n+m)}{m}\pa,\no\\
\left(J_2^{(n,m)}\right)_{i,n-1}&=&
\frac{n(i+m+1)}{m}v_{i+1}\pa,\no\\
\left(J_2^{(n,m)}\right)_{n-1,j}&=&\frac{n(j+m+1)}{m}\pa
v_{j+1},\no\\ \left(J_2^{(n,m)}\right)_{i,j}&=&
(n-i-1)v_{i+j+2-n}\pa+(n-j-1)\pa v_{i+j+2-n} \no\\
&&+\sum_{k=j+2}^{n-1}[(k-i-1)v_{i+j+2-k}\pa v_k+(k-j-1)v_k\pa
v_{i+j+2-k}]\no\\ &&+\frac{(i+m+1)(j+1)}{m}v_{i+1}\pa v_{j+1}
 \eea where
$i,j=-m,-m+1,\cdots,n-2$ and $v_{i<-m}=0$. We refer the above
Poisson algebra to $w^{(n,m)}$-algebra.

For the Poisson algebra associated with the first structure, it
can be directly obtained from the Hamiltonian map (\ref{ham1}) for
the case of $m>0$ as follows:
 \[\label{positive}
  \left(J_1^{(n,m>0)}\right)_{ij}=
  \left\{
  \ba{ll}
    (i+1)v_{i+j+2}\pa+(j+1)\pa v_{i+j+2}, & -1\leq i,j\leq n-1 \\
    -(i+1)v_{i+j+2}\pa-(j+1)\pa v_{i+j+2}, & -m\leq i,j\leq -2 \\
    0 & \mbox{otherwise}.
  \ea
  \right.
 \]
However the case for $m<0$ requires the Dirac reduction and turns
out to be
 \[\label{negative}
  \left(J_1^{(n,m<0)}\right)_{ij}=\left(J_1^{(n,1)}\right)_{ij}-
  \sum_{k,l=-1}^{|m|-1}\left(J_1^{(n,1)}\right)_{ik}
  \left(J_1^{(n,1)}\right)^{-1}_{kl}\left(J_1^{(n,1)}\right)_{lj},
  \qquad |m|\leq i,j \leq n-1.
 \]
 Note that the bi-Hamiltonian structures of constrained
dmKP hierarchies can be cast into the following recursive formula:
 \[
 \left(J_1^{(n,m)}\right)_{ij}\frac{\de H_{k+n}}{\de v_j}=
 \left(J_2^{(n,m)}\right)_{ij}\frac{\de H_{k}}{\de v_j}.
 \]
Next, let us focus on the algebraic structures of the
$w^{(n,m)}$-algebra (\ref{poal2}). The first few of them are
 \bea
 \{v_{n-1}(x), v_{n-1}(y)\}_2^{(n,m)}&=& -\frac{n(n+m)}{m}\pa\cdot\de(x-y),\no\\
  \{v_{n-1}(x), v_{n-2}(y)\}_2^{(n,m)}&=& -\frac{n(n+m-1)}{m}\pa v_{n-1}(x)\cdot\de(x-y),
  \label{preag}\\
  \{v_{n-2}(x), v_{n-2}(y)\}_2^{(n,m)}&=& -\left[v_{n-2}(x)\pa+\pa v_{n-2}(x)+
  \frac{(n-1)(n+m-1)}{m}v_{n-1}(x)\pa v_{n-1}(x)\right]\cdot\de(x-y).\no
 \eea
In spite of the fact that $v_{n-1}$ satisfies the {\it
$U(1)$-Kac-Moody\/} algebra, the algebraic  structure shown above
is still unclear. However, if we define
 \be\label{virasoro}
 w_2(x)=v_{n-2}(x)-\frac{n-1}{2n}v^2_{n-1}(x)
 \ee
then the second and the third equations in (\ref{preag}) can be
rewritten as
 \bea
 \{v_{n-1}(x), w_2(y)\}_2^{(n,m)}&=& -[v_{n-1}(x)\pa+v_{n-1}'(x)]\cdot\de(x-y),\no\\
  \{w_2(x), w_2(y)\}_2^{(n,m)}&=& -[2w_2(x)\pa+w'_2(x)]\cdot
  \de(x-y)\no
 \eea
 where $w_2$, being a generator, is a Diff$S^1$ tensor with weight 2
  and  $v_{n-1}$ a tensor of weight 1. In fact, using (\ref{poal2}) and (\ref{virasoro}) we
  have
 \[
\{v_{n-i}(x), w_2(y)\}_2^{(n,m)}=
-[iv_{n-i}(x)\pa+v_{n-i}'(x)]\cdot\de(x-y)
 \]
 that means, except $v_{n-2}$, each coefficient $v_{n-i}$ in the Lax
operator $K_{(n,m)}$ , with respect to the generator $w_2$, is
already a Diff$S^1$ tensor with weight $i$. Hence the Poisson
algebra $w^{(n,m)}$ defined in (\ref{poal2}) is isomorphic to
$w_{(n+m)}$-$U(1)$-$Kac$-$Moody$-algebra generated by the primary
fields
 \[
w_1\equiv v_{n-1},\quad w_2\equiv
v_{n-2}-\frac{n-1}{2n}v_{n-1}^2,\quad w_i\equiv v_{n-i},
\quad(3\leq i\leq n+m)
 \]
 Note that the Diff$S^1$ flows can be viewed as the Hamiltonian
 flows generated by the Hamiltonian $\int \epsilon(x)h_1$ due to the
 fact that  $h_1=n\res(K_{(n,m)})^{1/n}=w_2$. That is the reason why the
 $w$-algebraic structure is encoded in the constrained dmKP hierarchy.
  Finally when we take the limit $m\to \infty$, (\ref{virasoro}) still holds
 and the Poisson algebra (\ref{poal2}) recovers to
 $w^{(n,\infty)}\equiv w^{(n)}_{dmKP}$ defined by the Lax operator
 $K_n$.

\section{KW theorem and free-field realizations}

It has been shown \cite{FR,MR,CL} that the second GD structure
defined by (\ref{gd2}) has nice properties with respect to the
factorization of the associated Lax operator. For example, if we
factorize $L=L_1L_2$, then
 \be\lb{kw1}
\{F, G\}_2^{GD}(L)=\{F, G\}_2^{GD}(L_1)+\{F, G\}_2^{GD}(L_2).
 \ee
On the other hand, if $L=L_1^{\alpha},  \alpha \in N$ then we have
 \be\lb{kw2}
\{F, G\}_2^{GD}(L)=\frac{1}{\alpha}\{F, G\}_2^{GD}(L_1).
 \ee
Eqs.(\ref{kw1}) and (\ref{kw2}) are just the corresponding KW
theorem for the classical limit of the second GD bracket. More
generally, we shall consider the factorization of the polynomial
$L_{n+m}$ of the form
 \be\lb{fact}
L_{n+m}=\prod_{i=1}^{l}(p+\phi_i)^{\alpha_i}, \qquad
\sum_{i=1}^{l}\alpha_i=n+m,
 \ee
  where the Miura variables $\phi_i$ are zeros of
$L_{n+m}$ with multiplicities $\alpha_i$, then (\ref{kw1}) and
(\ref{kw2}) imply that \be\label{mgd2} \{F,
G\}^{GD}_2(L_{n+m})=-\sum_{i=1}^{l}\frac{1}{\alpha_i}\int
\left(\frac{\de F}{\de \phi_i}\right)' \left(\frac{\de G}{\de
\phi_i}\right).
 \ee
 To complete the discussion of the KW theorem we have to treat the
third GD structure under the factorization (\ref{fact}). In fact,
we can show that the third GD structure enjoys the following
property [for the proof, see Appendix A]:
 \be\lb{kw3}
\{F, G\}_3^{GD}(L_1^{\alpha}L_2)=\{F, G\}_3^{GD}(L_1L_2).
 \ee
 That means the multiplicities $\alpha_i$ do not involve in
the KW theorem with respect to the third GD structure.
Hence,
 \bea\label{mgd3}
\{F, G\}^{GD}_3(L_{n+m})&=& \{F,
G\}^{GD}_3(\prod_{i=1}^{l}(p+\phi_i)),\no\\
&=&\sum_{i,j=1}^{l}\int \left(\frac{\de F}{\de \phi_i}\right)'
\left(\frac{\de G}{\de \phi_j}\right).
 \eea
Combining (\ref{mgd2}) and (\ref{mgd3}) we have
\[\label{kw}
  \{F, G\}^{GD}_{2+3}(L_{n+m})=
  -\sum_{i,j=1}^{l}\left(\frac{1}{\alpha_i}\de_{ij}-\frac{1}{m}\right)
  \int\left(\frac{\de F}{\de \phi_i}\right)' \left(\frac{\de G}{\de \phi_j}\right).
 \]
Among other things, the fundamental brackets for the Miura
variables $\phi_i$ are
 \bea
  \{\phi_i(x), \phi_j(y)\}^{(n,m)}_2(K_{(n,m)})&=&
  \{\phi_i(x), \phi_j(y)\}^{GD}_{2+3}(L_{n+m})\no\\
  &=&\left(\frac{1}{\alpha_i}\de_{ij}-\frac{1}{m}\right)\pa\cdot\de(x-y)
  \label{miubk}
 \eea
 where $i,j=1,2,...,l$.

  Since the above Poisson matrix is symmetric and hence can be
diagonalized by linearly combining the Miura variables $\phi_i$ to
obtain the free fields. For example, suppose all zeros are simple,
i.e. $\alpha_i=1, i=1,2,\cdots, n+m$, then $v_i=S_{n-i}(\phi_j)$
being the symmetric functions of $\{\phi_j\}$ and the Poisson
matrix (\ref{miubk}) becomes
$P_{2+3}={\mathbf{1}}-\frac{1}{m}{\mathbf{h}}{\mathbf{h}}^T$ where
$T$ denotes the transpose operation, $\mathbf{1}$ is a
$(n+m)\times (n+m)$ identity matrix and
${\mathbf{h}}^T=(1,\cdots,1)$. It is quite easy to construct
$n+m-1$ orthonormal eigenvectors ${\mathbf{h}}_i$ as follows
 \bea
{\mathbf{h}}^T_1&=&(1,-1,0\cdots,0)/\sqrt{2},\no\\
{\mathbf{h}}^T_2&=&(1,1,-2\cdots,0)/\sqrt{6},\no\\
 &\cdots&\no\\
{\mathbf{h}}^T_{n+m-1}&=&(1,1,\cdots,1,-n-m+1)/\sqrt{(n+m)(n+m-1)}\no
 \eea
which satisfy $P_{2+3}{\mathbf{h}}_i={\mathbf{h}}_i$ and hence
have eigenvalue $+1$. Finally, from orthogonality, the remaining
orthonormal eigenvector has the form
 \[
{\mathbf{h}}^T_{n+m}=(1,1,\cdots,1)/\sqrt{n+m}
 \]
with eigenvalue $-n/m$. Now if we rewrite the Miura variables
${\mathbf{\phi}}^T=(\phi_1,\phi_2,\cdots,\phi_{n+m})$ as $
\phi=\mathbf{He}$
 where $\mathbf{H}$ is a $(n+m)\times(n+m)$ matrix defined by
${\mathbf{H}}=[{\mathbf{h}}_1,{\mathbf{h}}_2,\cdots,{\mathbf{h}}_{n+m}]$,
then
 \be\label{free}
\{e_i(x), e_j(y)\}_2^{(n,m)}=\la_i\pa\cdot\de(x-y)
 \ee
 with $\la_i=1$ $(i=1,2,\cdots,n+m-1)$, $\la_{n+m}=-n/m$. Therefore (\ref{free})
 provides a free-field realization of the $w^{(n,m)}$-algebras (\ref{poal2}) and
the Lax operator $K_{(n,m)}$ can be expressed as
 \[
K_{(n,m)}=\prod_{i=1}^{n+m}(p+({\mathbf{He}})_i)p^{-m},
 \]
where the free fields $e_i$ satisfy the Hamiltonian flows
 \[\label{freefl}
  \frac{\pa e_i}{\pa t_k}=-\la_i\left(\frac{\de H_k}{\de
  e_i}\right)'.
 \]
In the case of $m\to \infty$, the Poisson matrix of (\ref{miubk})
becomes diagonal, which provides the free-field realization of the
$w^{(n)}_{dmKP}$-algebra.

\section{Examples}

{\bf Example 1 :\/} For the Lax operator
$K_{(2,1)}=p^2+v_1p+v_0+v_{-1}p^{-1}$ the first nontrivial
equations  are given by \bea \frac{d}{dt_2}\left( \ba{c} v_1
\\ v_0\\ v_{-1} \ea \right) &=& \left( \ba{c} 2v_{0x}\\
v_1v_{0x}+2v_{-1x}\\ (v_1v_{-1})_x \ea \right),\no\\
8\frac{d}{dt_3} \left( \ba{c} v_1 \\ v_0\\ v_{-1} \ea \right) &=&
\left( \ba{c} -3v_1^2v_{1x}+12(v_0v_1)_x+24v_{-1x}\\
12v_{1x}v_{-1}+24v_1v_{-1x}+12v_0v_{0x}+3v_1^2v_{0x}\\
12(v_0v_{-1})_x+3(v_1^2v_{-1})_x \ea \right)\no
 \eea
  which are first equations of the generalized Benney hierarchy. The first
Hamiltonians of these hierarchy flows are given by
 \bea H_1&=&\int
\left(v_0-\frac{1}{4}v_1^2\right), \no\\ H_2&=&\int v_{-1}, \no\\
H_3&=&\int \left( \frac{1}{2}v_1v_{-1}+\frac{1}{4}v_0^2-
\frac{1}{8}v_0v_1^2+\frac{1}{64}v_1^4\right), \no\\ H_4&=&\int
v_0v_{-1},\no\\ H_5&=&\int
\left(-\frac{1}{512}v_1^6+\frac{3}{128}v_1^4v_0-\frac{1}{16}v_1^3v_{-1}
-\frac{3}{32}v_1^2v_0^2+\frac{3}{4}v_1v_0v_{-1}+\frac{1}{8}v_0^3+\frac{3}{4}v_{-1}^2\right).\no
\eea
 Then the Lax flows can be rewritten as Hamiltonian flows as
follows:
 \bea
 \frac{d}{dt_2}{\mathbf{v}}&=&J_1^{(2,1)}\frac{\de H_4}{\de{\mathbf{v}}}
 =J_2^{(2,1)}\frac{\de H_2}{\de{\mathbf{v}}}\no\\
 \frac{d}{dt_3}{\mathbf{v}}&=&J_1^{(2,1)}\frac{\de H_5}{\de{\mathbf{v}}}
 =J_2^{(2,1)}\frac{\de H_3}{\de{\mathbf{v}}}\no
 \eea
  where ${\mathbf{v}}^T=(v_1,v_0,v_{-1})$,
  $(\de H_i/\de {\mathbf{v}})^T=(\de H_i/\de v_1, \de H_i/\de v_0, \de H_i/\de v_{-1})$ and
   \bea J_1^{(2,1)}&=& \left( \ba{ccc} 0 & 0 &
2\pa \\ 0 & 2\pa & v_1\pa\\ 2\pa & \pa v_1 & 0 \ea \right),\no\\
J_2^{(2,1)}&=& \left( \ba{ccc} 6\pa & 4\pa v_1 & 2\pa v_0 \\
4v_1\pa & v_0\pa+\pa v_0+2v_1\pa v_1 & 2\pa
v_{-1}+v_{-1}\pa+v_1\pa v_0 \\ 2v_0\pa & \pa
v_{-1}+2v_{-1}\pa+v_0\pa v_1 & \pa v_1v_{-1}+v_1v_{-1}\pa \ea
\right).\no
 \eea
 On the other hand, the Lax operator $K_{(2,1)}$ can be expressed
 in terms of primary fields as
 \[
 K_{(2,1)}=p^2+w_1p+(w_2+\frac{1}{4}w_1^2)+w_3p^{-1}
 \]
where $w_i$ satisfy the {\it $w_3$-U(1)-Kac-Moody\/}-algebra
 \bea
 \{w_1(x), w_1(y)\}_2^{(2,1)}&=&-6\pa\cdot\de(x-y),\no\\
 \{w_1(x), w_2(y)\}_2^{(2,1)}&=&-[w_1(x)\pa+w_1'(x)]\cdot\de(x-y),\no\\
 \{w_1(x), w_3(y)\}_2^{(2,1)}&=&-[(2w_2(x)+\frac{1}{2}w_1^2(x))\pa+(2w_2(x)+w_1^2(x))']
 \cdot\de(x-y),\no\\
 \{w_2(x), w_2(y)\}_2^{(2,1)}&=&-[2w_2(x)\pa+w_2'(x)]\cdot\de(x-y),\no\\
 \{w_3(x), w_2(y)\}_2^{(2,1)}&=&-[3w_3(x)\pa+w_3'(x)]\cdot\de(x-y),\no\\
 \{w_3(x), w_3(y)\}_2^{(2,1)}&=&-[2w_1(x)w_3(x)\pa+(w_1(x)w_3(x))']\cdot\de(x-y).\no
 \eea
The free-field realization of the above algebra is given by
 \bea
 w_1&=&\sqrt{3}e_3,\no\\
 w_2&=&e_3^2-\frac{1}{2}e_2^2-\frac{1}{2}e_1^2,\no\\
 w_3&=&-\frac{1}{2\sqrt{3}}(e_1^2+e_2^2)e_3+
 \frac{1}{\sqrt{6}}(e_1^2-\frac{1}{3}e_2^2)e_2+\frac{1}{3\sqrt{3}}e_3^3\no
 \eea
with
 \bea
 \{e_1(x),e_1(y)\}_2^{(2,1)}&=&\{e_2(x),e_2(y)\}_2^{(2,1)}=\pa\cdot\de(x-y),\no\\
 \{e_3(x),e_3(y)\}_2^{(2,1)}&=&-2\pa\cdot\de(x-y).\no
 \eea

 {\bf Example 2 :\/} For the Lax operator
$K_{(3,-1)}=p^3+v_2p^2+v_1p$, the first nontrivial Lax  equations
are \bea 3\frac{d}{dt_2}\left( \ba{c} v_2 \\ v_1 \ea \right) &=&
\left( \ba{c} 6v_{1x}-2v_2v_{2x}\\ 2v_2v_{1x}-2v_{2x}v_1 \ea
\right),\no\\ 81\frac{d}{dt_4} \left( \ba{c} v_2 \\ v_1 \ea
\right) &=& \left( \ba{c} (5v_2^4-36v_2^2v_1+54v_1^2)_x\\
-4v_2^3v_{1x}+12v_2^2v_{2x}v_1-36v_{2x}v_1^2 \ea \right)\no
 \eea
 which are first equations of the dispersionless modified KdV hierarchy.
 The Hamiltonian flows are defined by
 \bea
  \frac{d}{dt_2}{\mathbf{v}}&=&J_1^{(3,-1)}\frac{\de
H_5}{\de{\mathbf{v}}}
 =J_2^{(3,-1)}\frac{\de H_2}{\de{\mathbf{v}}}\no\\
 \frac{d}{dt_3}{\mathbf{v}}&=&J_1^{(3,-1)}\frac{\de H_7}{\de{\mathbf{v}}}
 =J_2^{(3,-1)}\frac{\de H_4}{\de{\mathbf{v}}}\no
 \eea
with the first Hamiltonians
 \bea
 H_1&=&\int \left(v_1-\frac{1}{3}v_2^2 \right),\no\\
  H_2&=&\int
\left(\frac{2}{27}v_2^3-\frac{1}{3}v_2v_1\right),\no\\ H_4&=&\int
\left(-\frac{2}{243}v_2^5+\frac{5}{81}v_2^3v_1-\frac{1}{9}v_2v_1^2\right),
\no\\ H_5&=&\int
\left(\frac{7}{2187}v_2^6-\frac{7}{243}v_2^4v_1+\frac{2}{27}v_2^2v_1^2
-\frac{1}{27}v_1^3\right), \no\\ H_7&=&\int
\left(-\frac{11}{19683}v_2^8+\frac{44}{6561}v_2^6v_1
-\frac{20}{729}v_2^4v_1^2+\frac{10}{243}v_2^2v_1^3-\frac{1}{81}v_1^4\right)
\no
 \eea
 and
  \bea J_1^{(3,-1)}&=& \left( \ba{cc}
9v_2v_1^{-1}\pa v_1^{-1}+9v_1^{-1}\pa v_2v_1^{-1} &
-9v_1^{-1}\pa+6v_2v_1^{-1}\pa v_2v_1^{-1}+6v_1^{-1}\pa
v_2^2v_1^{-1} \\ -9\pa v_1^{-1}+6v_2v_1^{-1}\pa
v_2v_1^{-1}+6v_2^2v_1^{-1}\pa v_1^{-1}  & -6\pa
v_2v_1^{-1}-6v_2v_1^{-1}\pa+8v_2v_1^{-1}\pa v_2v_1^{-1} \ea
\right),\no\\ J_2^{(3,-1)}&=& \left( \ba{cc} -6\pa & -3\pa v_2 \\
-3v_2\pa & v_1\pa+\pa v_1-2v_2\pa v_2 \ea \right).\no
 \eea
Rewriting the Lax operator $K_{(3,-1)}$ in terms of $w_i$ yields
 \[
 K_{(3,-1)}=p^3+w_1p^2+(w_2+\frac{1}{3}w_1^2)p
 \]
where $w_1$ and $w_2$ satisfy the (centerless-){\it
Virasoro-U(1)-Kac-Moody\/} algebra, namely,
 \bea
\{w_1(x),w_1(y)\}_2^{(3,-1)}&=&6\pa\cdot\de(x-y),\no\\
\{w_1(x),w_2(y)\}_2^{(3,-1)}&=&-[w_1(x)\pa+w_1'(x)]\cdot\de(x-y),\no\\
\{w_2(x),w_2(y)\}_2^{(3,-1)}&=&-[2w_2(x)\pa+w_2'(x)]\cdot\de(x-y).\no
 \eea
The free-field realization of the above algebra can be easily
obtained as
 \bea
 w_1&=&\sqrt{2}e_2,\no\\
 w_2&=&-\frac{1}{2}e_1^2-\frac{1}{6}e_2^2\no
 \eea
with
 \bea \{e_1(x),e_1(y)\}_2^{(3,-1)}&=&\pa\cdot\de(x-y),\no\\
\{e_2(x),e_2(y)\}_2^{(3,-1)}&=&3\pa\cdot\de(x-y).\no
 \eea


\section{Concluding remarks}

We have studied the constrained dmKP hierarchies from the dmKP
hierarchy by truncating the Lax operator $K_n$ to any finite
order. We have obtained the compatible bi-Hamiltonian structures
of constrained dmKP hierarchies via the Dirac reduction and
written down their associated Poisson algebras explicitly. We show
that the second Poisson algebra $w^{(n,m)}$ turns out to be the
$w_{(n+m)}$-$U(1)$-$Kac$-$Moody$-algebra. Its free-field
realization can be obtained via the corresponding KW theorem.
Several examples including the generalized Benney hierarchy have
been used to illustrate the obtained results.

We would like to remark that the bi-Hamiltonian structures
obtained in this paper are of hydrodynamic type \cite{ND}, i.e.
the Hamiltonian operators can be expressed as [for convention, we
have to rewrite the Hamiltonian operators as contravariant tensors
]
 \[\label{hydro}
J^{ij}(v)=g^{ij}(v)\pa-\Ga^{ij}_k(v)v^k_x
 \]
 where, under the non-degenerate condition $\det(g_{ij}(v))\neq 0$,
$g_{ij}(v)\equiv (g^{ij})^{-1}$ can be viewed as a (pseudo-)
Riemannian metric and $\Ga^k_{ij}(v)\equiv g_{il}\Ga^{kl}_j$ are
the components of the Levi$-$Civit\`a connection defined by
$g_{ij}(v)$. Moreover, the Jacobi identity of the Hamiltonian
structures implies that the metric is flat. This can be easily
checked for the illustrated  examples. On the other hand, it was
pointed out \cite{Li} that the third (or higher) Hamiltonian
structures may induce non-local Hamiltonian operators which also
possess nontrivial geometrical interpretations \cite{F} and thus
deserve more investigations.


{\bf Acknowledgements\/}\\
 We would like to thank Prof. J.C Shaw
for helpful discussions. J.H.Chang thanks for the support of the
Academia Sinica and M.H.Tu  thanks for the support of  the
National Science Council of Taiwan under Grant No. NSC
89-2112-M194-018.\\

\appendix
\section{A proof of (4.5)}

Let $L=L_1^{\alpha}L_2$ then the variation
\[
\de F=\int \res\left(\de L\frac{\de F}{\de L}\right)= \int
\res\left(\de L_1\frac{\de F}{\de L_1}+\de L_2\frac{\de F}{\de
L_2}\right)
\]
gives the relations
\[
\frac{\de F}{\de L_1}=\alpha L_1^{\alpha-1}L_2\frac{\de F}{\de L},
\qquad \frac{\de F}{\de L_2}=L_1^{\alpha}\frac{\de F}{\de L}.
\]
Hence
 \bea
\{F, G\}_3^{GD}(L)
 &=& \int\res\left( \left[\left[ L_1^{\alpha}L_2,
 \int^x\res\left[\left[ L_1^{\alpha}L_2, \frac{\de F}{\de L}\right]\right]\right]\right]
 \frac{\de G}{\de L}\right),\no\\
 &=&\int\res\left( \left[\left[ L_1^{\alpha}L_2,
 \int^x\res\left(\left[\left[ L_1, \frac{\de F}{\de
 L_1}\right]\right]+\left[\left[ L_2, \frac{\de F}{\de
 L_2}\right]\right]\right)\right]\right] \frac{\de G}{\de L}\right),\no\\
 &=&\int\res\left(\left[\left[ L_1,
 \int^x\res\left(\left[\left[ L_1, \frac{\de F}{\de
 L_1}\right]\right]+\left[\left[ L_2, \frac{\de F}{\de
 L_2}\right]\right]\right)\right]\right] \frac{\de G}{\de L_1} \right)+(1\leftrightarrow 2).\no\\
 \eea
Now define $\hat{L}=L_1L_2$ then
\[
\frac{\de F}{\de L_1}=\frac{\de F}{\de \hat{L}}L_2, \qquad
\frac{\de F}{\de L_2}=\frac{\de F}{\de \hat{L}}L_1
\]
and
 \bea
\mbox{(A1)} &=&\int \res\left(L_2\left[\left[L_1,
\int^x\res\left[\left[\hat{L}, \frac{\de F}{\de
\hat{L}}\right]\right]\right]\right] \frac{\de G}{\de
\hat{L}}\right)
 +(1\leftrightarrow 2),\no\\
 &=&\int \res\left(\left[\left[\hat{L}, \int^x\res\left[\left[\hat{L}, \frac{\de F}
 {\de \hat{L}}\right]\right]\right]\right] \frac{\de G}{\de \hat{L}}\right),\no\\
 &=&\{F,G\}_3^{GD}(L_1L_2).\qquad \Box\no
 \eea



\begin{thebibliography}{99}

\bibitem{LM}
D. Lebedev and Yu. Manin, Phys. Lett. A {\bf 74\/}, 154 (1979).

\bibitem{TT} K. Takasaki and T. Takebe, Rev. Math. Phys. {\bf 7\/}, 743 (1995).

\bibitem{CT}
Jen-Hsu Chang and Ming-Hsien Tu, solv-int/9912016.

\bibitem{Li}
Luen-Chau Li, Commun. Math. Phys. {\bf 203\/}, 573 (1999).

\bibitem{FR}
J.M. Figueroa-O'Farrill and E. Ramos, Phys. Lett. B {\bf 282\/},
357 (1992).

\bibitem{KO}
B. Konopelchenko and W. Oevel, Publ. RIMS, Kyoto Univ. {\bf 29\/},
581 (1993).

\bibitem{OS}
 W. Oevel and W. Strampp, Comm. Math. Phys. {\bf 157\/}, 51 (1993).

\bibitem{O}
W. Oevel, Phys. Lett. A {\bf 186\/}, 79 (1994).

\bibitem{MR}
J. Mas and E. Ramos, Phys. Lett. B {\bf 351\/}, 194 (1995).

\bibitem{CL}
Y. Cheng and Z.F. Li, Lett. Math. Phys. {\bf 42\/}, 73 (1997).

\bibitem{ND}
B. Dubrovin and S.P. Novikov, Russ. Math. Surv. {\bf 44\/}, 35
(1989).

\bibitem{F}
E.V. Ferapontov, in {\it Topics in Topology and Mathematical
Physics\/}, Amer. Math. Soc. Transl. Ser2. {\bf 170\/}, 33 (1995).


\end{thebibliography}
\end{document}